\documentstyle[12 pt]{article}
\newcommand\Journal[4]{{#1} {\bf #2}, #3 (#4)}

\newcommand\NPB{{\em Nucl. Phys.} B}
\newcommand\PLB{{\em Phys. Lett.}  B}
\newcommand\PRL{\em Phys. Rev. Lett.}
\newcommand\PRD{{\em Phys. Rev.} D}

\newcommand\EPJC{{\em Eur. Phys. J.} C}
\newcommand\IJMP{{\em Int. Jout. Mod. Phys.} }
\newcommand\APPB{{\em Acta. Phys. Polon.} B}
\newcommand\PTP{{\em Prog. Theor. Phys.} B}

\def\hybrid{\topmargin +10pt    \oddsidemargin 0pt
\headheight 0pt \headsep 0pt
       \textwidth 6.5in        
       \textheight 9in         
        \marginparwidth .875in \parskip 5pt plus 1pt   \jot = 1.5ex}
        \hybrid
        \newcommand{\PSbox}[3]{\mbox{\rule{0in}{#3}
\includegraphics{#1}\hspace{#2}}}
\begin{document}
\begin{flushright}
MIT-CTP-2911\\
hep-ph/yymmddd
\end{flushright}
\vspace{0.3in}

\begin{center}
{\bf  \LARGE Higgs Sector in Anomaly-Mediated 
Supersymmetry Breaking Scenario} \\
\vspace{0.4in}

{Shufang Su}\\
\vspace{.05in}
{ \small \it Center for Theoretical Physics\\
Laboratory for Nuclear Science and Department of Physics\\
Massachusetts Institute of Technology\\
Cambridge, MA 02139, USA\\}
\vspace{0.2 in} {\bf Abstract} \\
\end{center}
{\small  In the minimal anomaly-mediated supersymmetry breaking (AMSB)
model, a universal contribution  $m_{0}$  to all the scalar masses is
introduced  in order to avoid the negative slepton mass problem.   The
Higgs spectrum and couplings are determined by four parameters:
$m_{\rm aux},\ m_{0},\ \tan\beta$ and sign ($\mu$).  The sign of $\mu$
affects $m_A$  at large $\tan\beta$ and  $m_h$ at small $\tan\beta$.
The CP-odd Higgs mass  $m_A$ is usually much larger than ${m}_{Z}$ and
the lightest CP-even Higgs is simply analogous to the one in  the
standard model.   The current and future  Higgs searches in LEP,
Tevatron and LHC  provide a test ground for   the AMSB scenario.
The current LEP bounds and LEP 192/196 preliminary results  have
already excluded a small  $m_{0}$ and $m_{\rm aux}$ region for small
$\tan\beta$.  While the entire parameter space will be excluded if no
Higgs is found at Tevatron RUN II with 2  ${\rm fb}^{-1}$
luminosity. However, if the AMSB scenario is true,  a Higgs can be
found at 5$\sigma$ significance level at both Tevatron running at
luminosity 10 ${\rm fb}^{-1}$ or higher and LHC.}
\vspace{0.4 in}
\normalsize

\section{Introduction}
The mechanism for electroweak symmetry
breaking (EWSB) remains a mystery  after many years of effort on both
the theoretical and experimental sides.  In the standard model, the
Higgs mechanism can give mass to all the  known particles and is left
with a fundamental scalar boson.  The best fit  to the electroweak
precision measurements prefers a Higgs mass of $90-100$ GeV
\cite{higgsmass}.  However,  theoretically, there is no protection of
this  scalar mass and it can receive large radiative corrections 
from new physics at any high scale $\Lambda$ (e.g. the GUT scale).
It is thus a miracle that the cancellations
between the large scales should be  so complete  as to 
give the preferred small Higgs mass. 

Supersymmetry (SUSY) provides a good explanation of this hierarchy
problem  if the SUSY breaking scale is around 1 TeV. There  are two
Higgses  $H_{u}$ and $H_{d}$, which give masses to up type quarks  and
down type quarks/leptons respectively.  There are five Higgs bosons
left after the EWSB: two CP-even Higgses $h$ and $H$ with
$m_{h}<m_{H}$,  one CP-odd Higgs $A$ and two charged Higgses $H^+$ and
$H^-$.   Unlike the standard model, where basically no upper limit is put
on the Higgs mass  except the naturalness, the light CP-even Higgs
mass $m_{h}$  in the supersymmetric models is highly constrained.  The
tree level mass is   lighter than $m_{Z}$, while the radiative
corrections could change  the result by $20-30$ GeV.  Lots of
calculations have been done on the radiative corrections to $m_h$ 
using different approximations such as the diagrammatic 
approach \cite{dia,sven},  the renormalization
group approach \cite{renor} and the effective potential  method
\cite{effec}.  In this paper, we adopt the renormalization group
improved one-loop effective potential  method discussed in \cite{carena},
including the important next-to-leading effects.

The newly proposed Anomaly-Mediated Supersymmetry Breaking (AMSB) 
scenario \cite{lr,giudice} presented an alternate way to give mass to 
all the super particles. SUSY breaking happens on a separate brane and is 
communicated to the visible world via the super-Weyl anomaly. 
The overall scale of sparticle masses are set by $m_{\rm aux}$, 
which is the VEV of the auxiliary field in the supergravity multiplet.
In the minimal case,  as will be explained in detail in the 
next section,
the particle spectrum can be determined by 
$3+1$  parameters:  
\begin{equation}
\{m_{\rm aux},\ m_{0},\ \tan\beta,\ {\rm sign}(\mu).\}
\end{equation}
Here $m_{0}$ is some universal mass scale introduced at the GUT scale to 
solve the tachyonic slepton mass problem. 
 
Some phenomenological implications of AMSB have been studied in  the
literature.   There is a novel ``focus point'' behavior in $\mu$
\cite{feng,feng1},  which allows squark and slepton masses far above
their usual naturalness bounds.  Sleptons are nearly degenerate and
highly mixed, while the squarks are universally  very heavy
{\cite{feng,wells,kribs}.   The experimental signatures are sensitive to the
hierarchy of sneutrino,  slepton and Wino masses.  In particular, the
neutral Wino as the Lightest Supersymmetric Particle (LSP) scenario
has been explored in detail in  \cite{winolsp}.  A disappearing 
charged track inside the detector serves as a distinctive experimental 
signal.  A
variety of low energy observables, including $b\rightarrow{s}\gamma$,
the anomalous magnetic moment of the muon and the electric dipole
moments of the electron and neutron, are sensitive probes of
anomaly-mediated parameter spaces \cite{feng}.  AMSB scenario has
important   cosmological consequences as well \cite{wells, takeo}.  The
gravitino mass is much heavier than the mass of the other sparticles, 
which alleviate the cosmological
problem associated with gravitino decays during nucleosynthesis.
Furthermore, the neutral Wino-LSP can form the cold dark matter, because
they are copiously produced from the primordial gravitino decays. 
However, no detailed presentation of the light Higgs phenomenology in 
the AMSB scenario has existed so far.  It is thus necessary to study the 
Higgs sector and see how the current or the future experimental Higgs 
searches can constrain the AMSB parameter spaces. 

The current experimental Higgs bound $m_{H_{SM}}>95.2$ GeV\cite{current}
is set by the LEP combined searches at 189 GeV through the channel
$e^+e^-\rightarrow{Z}{H_{SM}}$ where $H_{SM}\rightarrow{b}\bar{b}\ {\rm
or}\ \tau\bar\tau$.  A recent LEP updates at 192/196 GeV and 
109 ${\rm pb}^{-1}$ luminosity gave a preliminary limit of 
98.7 GeV \cite{current} on the standard model Higgs mass. 
After the final run of
LEP at $\sqrt{s}=200$ GeV, a lower limit of 108 GeV on the Higgs  
mass can be achieved if no Higgs is found \cite{lep}.  In
LHC \cite{LHC}, $pp\rightarrow\gamma\gamma+X$ has been studied.  For a
low luminosity of 30 ${\rm fb}^{-1}$, a standard  model Higgs in the  mass
range $105-148$ GeV can be discovered at  $5\sigma$ significance level.
With higher luminosity 100  ${\rm fb}^{-1}$, the reach can be extended
to $80-160$ GeV. If other decay channels
$H\rightarrow{ZZ},{ZZ^*}\rightarrow{4}l$,
$H\rightarrow{ZZ},{WW}\rightarrow{2l2\nu}$,
$H\rightarrow{WW}\rightarrow{l\nu{2}j}$  are included,  a reach of
600 GeV  or even higher is possible. 

In Tevatron RUN II and RUN III, the upgraded {CDF/D\O} will have greatly
improved  sensitivity in searching for the Higgs boson in both the
standard and supersymmetric models\cite{conway}. For light standard model Higgs
($m_{H}<135$ GeV),   the dominant Higgs decay mode is
$H\rightarrow{b}{\bar{b}}$.  Four search channels have been studied:
$l\nu{b}\bar{b}$ ($WH$ events where  $W\rightarrow{l}\nu$),
$\nu\bar\nu{b}\bar{b}$ ($ZH$ events where
$Z\rightarrow{\nu}\bar\nu$), $l^+l^-b\bar{b}$ ($ZH$ events where
$Z\rightarrow{l^+}l^-$) and $q\bar{q}b\bar{b}$ ($WH$ and $ZH$ events
with  $W$ and $Z$ decaying to $q\bar{q}$ pairs).  For heavier Higgs,
$H\rightarrow{WW}$ dominates.  A single Higgs production via  gluon
fusion  gives $l^+l^-\nu\bar\nu$ final states.  Higgs can also  be
produced in conjunction with a vector boson, where the search channels can
be  $l^\pm{l}^\pm{j}j$ or $l^\pm{l}^{'\pm}l^\pm$.  Combining all the
search  channels from both experiments,  the integrated luminosity
needed to exclude the SM Higgs at $95\%$ CL, or discover it at
3$\sigma$ or 5$\sigma$ level of significance, is given in
\cite{conway}.  In this paper, we  use the results obtained by a
neural-network-based  analysis in separating the signals and background
for each hypothesized Higgs mass.

The light SUSY CP-even Higgs mass  is usually smaller than 130
GeV. Thus,  the strategy for light standard model Higgs  searches applies. The
exclusion/discovery reaches of SUSY Higgses  depend on both their masses
and couplings to standard model gauge bosons, quarks and leptons.  To
parameterize it in a model independent way, $R$ is introduced as 
\begin{eqnarray}
R&=&\frac{\sigma(p\bar{p}\rightarrow{W}h({\rm or}\ H))}
{\sigma(p\bar{p}\rightarrow{W}H_{SM})} \frac{B(h({\rm or}\ 
H)\rightarrow{b}\bar{b})} {B(H_{SM}\rightarrow{b}\bar{b})},\nonumber\\
&=&\sin^2(\beta-\alpha)\ ({\rm or}\ \cos^2(\beta-\alpha)) \frac{B(h({\rm
or}\ H)\rightarrow{b}\bar{b})} {B(H_{SM}\rightarrow{b}\bar{b})}.
\end{eqnarray}
In \cite{conway}, the exclusion/discovery contours of a light CP-even
Higgs in the space of $R$ and $m_{h}$ can be found for different
integrated luminosities, combining both of the {CDF/D\O} results. These contours
can be  translated to the reaches in the SUSY parameter spaces for
different  SUSY scenarios. 

The Higgs searches in the minimal Supersymmetric extension of the
Standard Model (MSSM) scenario have been discussed in detail
\cite{mssmsearch} for  LEP, Tevatron and
LHC.  Although certain regions of the parameter spaces are difficult
for some of the colliders, three collider experiments are complementary in
searching for light Higgs bosons.    The minimal Supergravity model
(mSUGRA) and Gauge Mediated SUSY Breaking model (GMSB)
\cite{othersearch} have also  been examined  for both Tevatron and LEP
2 Higgs searches.  Almost the entire parameter spaces can be covered by various
searching channels and stringent bounds can be put on these models
once the proposed signals are not found. 

The purpose of this paper is to work out the Higgs mass spectrum and
the couplings to the standard model particles in the AMSB  scenario.
Combining the existing results of the Higgs searches,  we can further
constrain the AMSB  parameter spaces.  In  Section~\ref{calculations},
we will discuss the AMSB scenario in detail and see how the entire SUSY
mass spectrum and Higgs couplings can be determined by the four
parameters.  In Section~\ref{results}, we presented the numerical
results for the Higgs masses in the parameter spaces of the AMSB scenario
and discuss their implications.  The discovery/exclusion regions for
Higgs searches in the AMSB parameter  spaces are also  presented, given
the existing Higgs search results in the literature.   We reserve the
last section for conclusions and discussions. 

\section{Calculations}
\label{calculations}
In the AMSB scenario, the low energy soft supersymmetry 
breaking parameters $M_i$ (gaugino masses, $i$=1$-$3),  
$m_{\rm scalar}^2$  and $A_{y}$ at the GUT scale are given by \cite{lr,wells}
\begin{eqnarray}
M_{\rm i}&=&\frac{\beta_{g_i}}{g_{\rm i}}m_{\rm aux},
\label{m1}\\
m_{\rm scalar}^2&=&-\frac{1}{4}\left(\frac{\partial\gamma}{\partial{g}}\beta_{g}+\frac{\partial\gamma}{\partial{y}}\beta_{y}\right)m_{\rm aux}^2+m_{0}^2,
\label{m2}\\
A_{y}&=&-\frac{\beta_{y}}{y}m_{\rm aux}.
\label{m3}
\end{eqnarray}
Notice that the slepton squared-masses would be negative if $m_{0}$ is absent. 
There have been several proposals to solve this tachyonic slepton 
problem: the bulk contributions \cite{lr}, the non-decoupling effects 
of ultra-heavy vectorlike matter fields \cite{negative}, 
coupling extra Higgs doublets to the leptons \cite{clm} and the heavy mass 
threshold contribution at higher orders \cite{kss}. 
Here we just adopted a phenomenological approach and introduced an additional 
mass scale $m_{0}$ at the GUT scale in order to keep the 
slepton masses positive \cite{wells}.  
For simplification, we choose $m_{0}$ to be the 
same for all the super scalar particles.  The deviation from this approach 
will be discussed in Section~\ref{conclusion}. 

Eq.~(\ref{m1}), (\ref{m2}) and (\ref{m3}) would be true at all 
scales if  $m_{0}$ is absent.  However, once $m_{0}$ is introduced
at the GUT scale,  the above definitions of $M_{i}$, $m_{\rm
scalar}^2$  and $A_{y}$ set the boundary conditions and
the entire SUSY spectrum can be obtained via the running of
supersymmetric Renormalization Group Equations (RGE) down to a lower
scale. 

Once we cross the squark  threshold, the squarks decouple and we are
left with an effective field  theory with two Higgses and all the
standard model particles\footnote{Gluinos  are also decoupled since their
masses are close to the squark masses.  The contributions of Bino and Winos to
the Higgs sector can be neglected   since the 
${\rm U(1)}_{Y}$ and SU(2) gauge couplings are small.}.  
The two unknown parameters $|\mu|$
and $b$ can be  determined by the minimization of the Higgs effective
potential  \cite{minimize} at the electroweak scale once we specify the
value of $\tan\beta$ and the sign of $\mu$.

The radiative corrections to the Higgs mass are important and can be as
large as $20-30$ GeV.  In this paper, we use the renormalization group
improved one-loop effective potential approach, including the
important  next-to-leading effects \cite{carena}.  One modification to 
\cite{carena, kribs} is that in the 
large  $\tan\beta$ case, the bottom mass corrections to the
bottom Yukawa coupling have been taken into account\cite{mssmsearch}:
\begin{equation}
 y_{b}=\frac{m_b}{v\cos\beta(1+\Delta(m_b))}.
\end{equation}
$\Delta(m_b)$ includes the one-loop contributions from gluino-sbottom
and higgsino-stop\cite{delta_mb}:
\begin{equation}
\Delta(m_b)\sim\frac{2\alpha_3}{3\pi}M_{\tilde{g}}\mu\tan\beta\
I(M_{\tilde{b}_1},M_{\tilde{b}_2},M_{\tilde{g}})+
\frac{Y_t}{4\pi}A_t\mu\tan\beta\
I(M_{\tilde{t}_1},M_{\tilde{t}_2},M_{\tilde{\mu}}),
\end{equation}
where $\alpha_3=g_3^2/4\pi$, $Y_t=y_t^2/4\pi$ and the function $I$ is
given by 
\begin{equation}
I(a,b,c)=\frac{a^2b^2\ln(a^2/b^2)+b^2c^2\ln(b^2/c^2)+c^2a^2\ln(c^2/a^2)}
{(a^2-b^2)(b^2-c^2)(a^2-c^2)}.
\end{equation}
In our conventions, $m_{\tilde{g}}$ and $A_t$ are both negative, while
$I$ is positive by definition.  Thus, for negative (positive)
$\mu$, the correction to bottom mass is positive (negative), which
decreases (increases) the bottom  Yukawa coupling.  This has an important
impact for $m_A$, as will be  shown in Section~\ref{results}. Also,
we choose to look at the Higgs mass  at the scale of on-shell top
quark mass $\bar{m}_{t}=m_t(m_t)$ because the  two-loop corrections
are small at that scale \cite{carena}.  

Therefore, the  Higgs 
mass spectrum  is fixed once we know the values of $m_{\rm aux}$, $m_{0}$ 
$\tan\beta$ and the sign of $\mu$. 
The couplings of Higgs particles to the other 
particles can be obtained by the diagonalization of the Higgs 
mass matrices.

\section{Results}
\label{results}
As was pointed out already in \cite{feng,feng1}, 
there is a `focus point' behavior 
of $\mu$ in the large $\tan\beta$ region. 
Though $\mu$ stays below 1 TeV, $m_{0}$ can be 
2 TeV or even larger.  So we choose our parameter ranges to be 
from 0 to 2 TeV for $m_{0}$, and from 20 TeV to 100 TeV for $m_{\rm aux}$, 
which corresponds to a wino mass 58 GeV $-$ 290 GeV. 

\begin{figure}
\PSbox{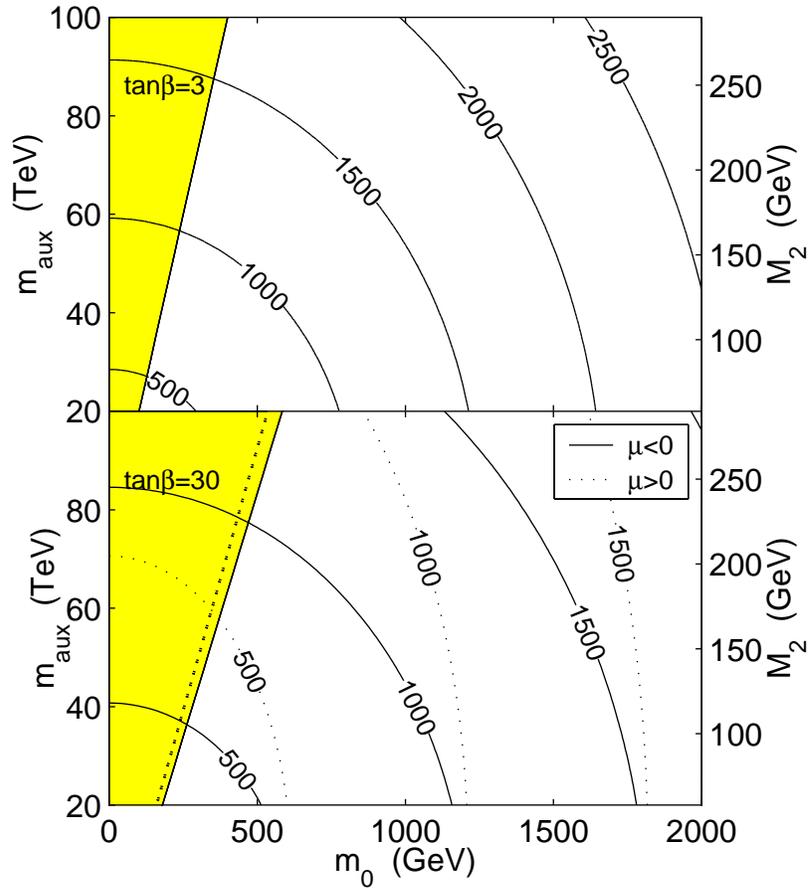 hoffset=-80 voffset=-250 hscale=100
vscale=100}{5.0in}{4.5in}
\caption{$m_{A}$ contours for $\tan\beta=3,30$.  For the upper plot, 
the results are the same for both signs of $\mu$.  The shaded region 
is excluded by the negative slepton masses.}
\label{mA}
\end{figure}

Figure~\ref{mA} shows the contours of the CP-odd Higgs mass $m_{A}$ in the
parameter space of $m_{0}$ and $m_{\rm aux}$, for two values of
$\tan\beta$=3, 30.   For small $\tan\beta$, changing the sign of $\mu$
does not change the value of $m_{A}$.  While for large $\tan\beta$,
$m_{A}$ decreases for positive $\mu$. This is because for
large $\tan\beta$, $m_{b}$ corrections  to the bottom Yukawa coupling
become important. Moreover, $y_{b}$ decreases (increases) for  negative
(positive) $\mu$ and $m_{A}$ is given by 
\begin{equation}
m_A^2=m_{H_d}^2+m_{H_u}^2+2\mu^2+{\rm radiative \ corrections}.
\end{equation}
The values of $m_{H_d}^2$ and ${m_{H_u}^2}$ at a lower scale are determined
by the RGE running from the GUT scale:
\begin{equation}
\frac{d}{dt}m_{i}^2\sim\frac{1}{16\pi^2}\left[
\sum_{j}Y^2m_j^2-g^2M_{\rm gaugino}^2+Y^2A^2\right],
\label{running}
\end{equation}
where the sum is over all the fields $\phi_j$ interacting with $\phi_i$ through
the Yukawa coupling $Y$.  Notice that the contributions from the terms 
proportional to the Yukawa coupling are always negative when running 
down from the GUT scale. 
In the AMSB scenario, $m_{H_d}^2>{m_{H_u}^2}$  
due to the large negative contributions to $m_{H_{u}}$ from the 
$y_t^2m_{j}^2$ terms.
Therefore, $m_A$ is  mainly determined by $m_{H_{d}}^2$, which is
larger for smaller $y_b$.   For small $\tan\beta$, $y_b$ is already
small and this effect becomes  less important. 

The shaded region in the upper-left corner is  excluded by the
negative slepton masses, which would disappear if we  drop the
assumption of the common mass $m_0$ for all the super scalar
particles, as will be discussed in Section~\ref{conclusion}.  In addition, for
large $\tan\beta$ and positive $\mu$,  small $m_{A}$ region (for
$\tan\beta$=30, $m_{A}<900$ GeV)  is already excluded  by
$b\rightarrow{s}\gamma$ constraints \cite{feng}. 

Notice that $m_{A}$ is usually larger than 500 GeV, which can
greatly simplify our analysis.  For $m_{A}\gg{m}_{Z}$, the low energy
effective field theory looks like the standard model with one Higgs boson
\begin{equation}
h\simeq{H}_{\rm SM}=H^0_1\cos\beta+H^0_2\sin\beta.
\end{equation}
In particular, the couplings of $hWW$, $hZZ$ and $hb\bar{b}$ are the
same as the standard model values.  All the results drawn from the
standard model Higgs can be directly applied to this light Higgs.  Most
relevantly,  the exclusion/discovery contours  in the AMSB parameter
spaces from the Higgs searches precisely coincide with the mass
contours of $m_{h}$.

Table~\ref{limit} summarizes the LEP, Tevatron and LEP reaches in the 
Higgs mass, either to exclude it at 95$\%$ CL or to discover it
at 5$\sigma$ level of significance\cite{current,lep,LHC,conway,mssmsearch}. 
\begin{table}
\begin{tabular}{c|c|c|c|c|c|c|c} \hline
&\multicolumn{3}{c|}{LEP}&\multicolumn{3}{c|}{Tevatron (RUN II, III)}
&LHC\\\cline{2-7}
&189 GeV&192/196 GeV&200 GeV&&&&\\
&current&109 ${\rm pb}^{-1}$&200 ${\rm pb}^{-1}$&
2 ${\rm fb}^{-1}$&10 ${\rm fb}^{-1}$&20 ${\rm fb}^{-1}$
&35 ${\rm fb}^{-1}$\\\hline
95$\%$ Exclusion&95.2&98.7&108&120&183.5&190&\\
5$\sigma$ Discovery&none&none&106.5&none&109.5&123&105-148\\\hline
\end{tabular}
\caption{95$\%$ CL exclusion/5$\sigma$ discovery limits on SM Higgs mass 
for LEP, Tevatron and LHC colliders.}
\label{limit}
\end{table}

\begin{figure}
\PSbox{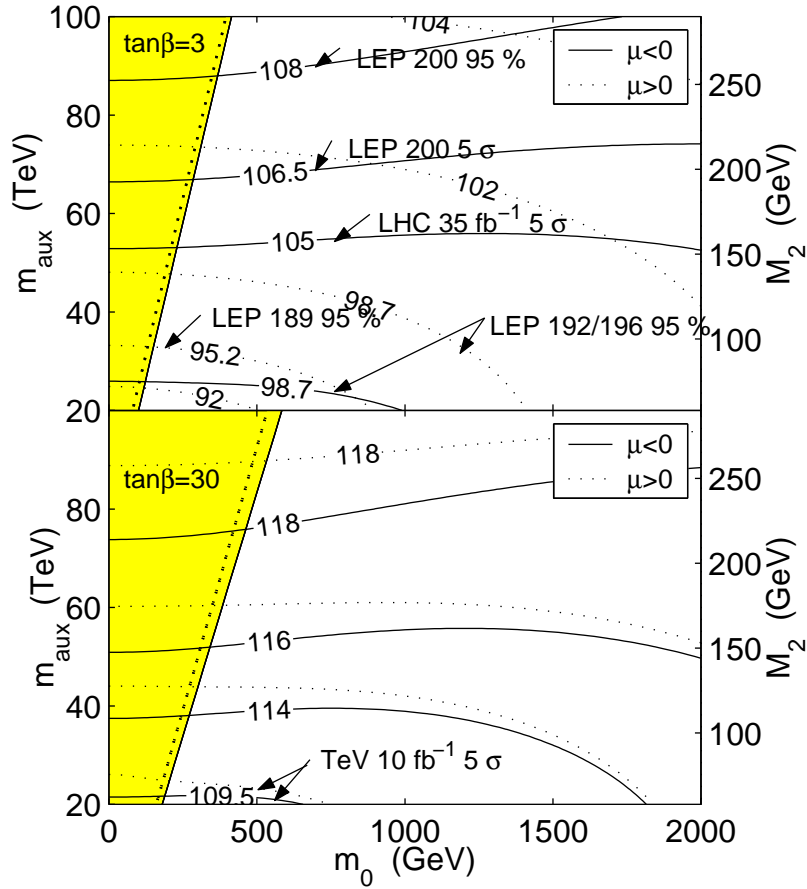 hoffset=-80 voffset=-250 hscale=100
vscale=100}{5.0in}{4in}
\caption{$m_{h}$ contours for $\tan\beta=3,30$. The shaded region 
is excluded by the negative slepton masses. The region below the 
mass contours would be excluded or discovered by different experiments, 
except for the LHC discovery contour, where the region above the 
contour would be discovered.}
\label{mh}
\end{figure}

For small $\tan\beta=3$,  $m_h$ is in
the range of 96 GeV $-$ 108 GeV  (92 $-$ 104 GeV) for $\mu<0$
($\mu>0$).  Current LEP bounds have already excluded a small $m_{0}$ and 
$m_{\rm aux}$ region for $\mu>0$.  The preliminary results of LEP running 
at 192/196  GeV can exclude the parameter region up to $m_{\rm aux}=47$ TeV 
($m_{\tilde{w}}=137$ GeV) and $m_{0}$=1440 GeV for $\mu>0$, while only a 
tiny region for $\mu<0$.  The entire region can be
excluded if  no Higgs is found after the LEP final run at 200 GeV.
However, if an AMSB Higgs does exist,  the first chance to discover it
is at LEP 200.  We can explore $m_{\rm aux}$ up to  74 TeV (corresponding to
$m_{\tilde{w}}=215$ GeV) with $\mu<0$.  
While for  $\mu>0$, the entire parameter space will be
covered.   Therefore, the AMSB scenario  can be examined for the
small $\tan\beta$ case in the next few years of  LEP
measurements. 

Tevatron experiments have great discovery potential  for
the Higgs boson. Even at the lowest integrated luminosity  2 ${\rm
fb}^{-1}$, the AMSB scenario will be excluded if no Higgs  is found.  For
luminosity 10 ${\rm fb}^{-1}$ or even higher, a CP-even  light Higgs
will be found in a natural parameter space of AMSB.   
The contour of LHC discovery should be read differently.  Only the region 
above the LHC 5$\sigma$ discovery contour will be covered by 
LHC Higgs searches at 30 ${\rm fb}^{-1}$.  This is because 
$pp\rightarrow\gamma\gamma+X$ is only sensitive to the Higgs mass in the range 
of 105 $-$ 148 GeV at that luminosity.  
Only part of the parameter space in $\mu<0$ case, nearly $m_{\rm aux}>53$ TeV 
($m_{\tilde{w}}>153$ GeV), will be discovered and $\mu>0$ will remain 
untouched.  Of course the entire parameter
space can be explored by LHC with luminosity 100 ${\rm fb}^{-1}$.   

The story is quite different for large $\tan\beta$.  The difference
in $m_h$ caused by the sign of $\mu$ is not so critical now.  This is
because $\mu$ only comes in through the combination
$\tilde{A}_t=A_t-\mu/\tan\beta$.  The  second term is much smaller at
large $\tan\beta$.  In average, $m_h$ is about 10 GeV larger than the
small $\tan\beta$ case, which eliminates its chance to be excluded or
discovered at LEP.  Tevatron running  at 2 ${\rm fb}^{-1}$ will
exclude the entire parameter space  if no light Higgs is found.
However, to discover a Higgs at 5$\sigma$  significance level, we
have to wait till Tevatron RUN III at  20 ${\rm fb}^{-1}$
integrated luminosity or the running of LHC. 

\begin{figure}
\PSbox{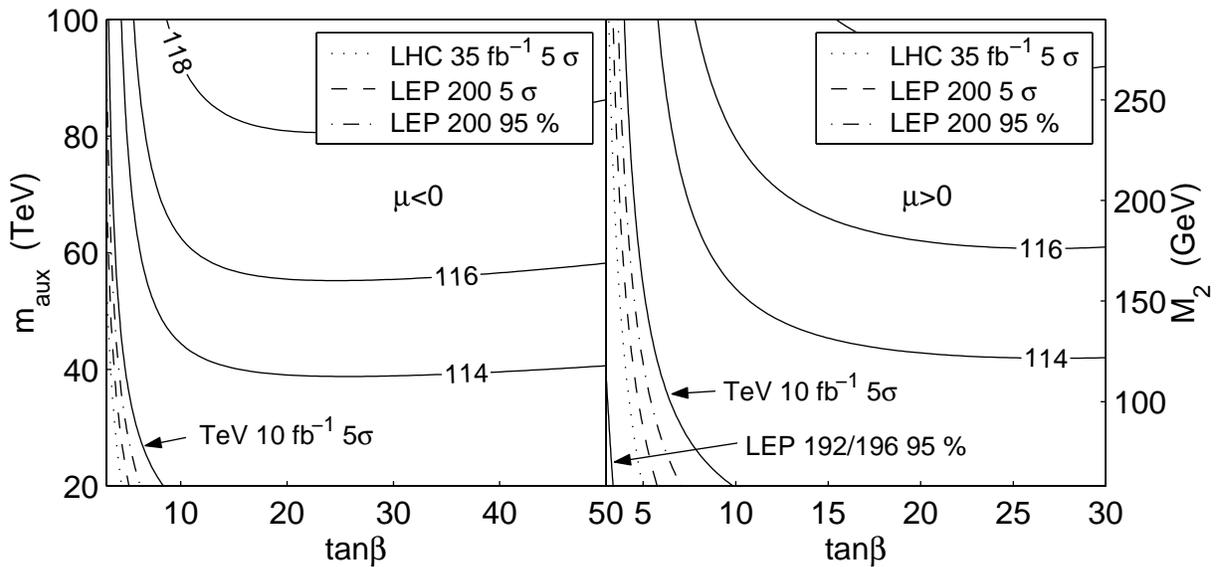 hoffset=-50 voffset=-200 hscale=100
vscale=100}{5.0in}{3in}
\caption{$m_{h}$ contours for $m_{0}=1000$ GeV. The region below the 
mass contours would be excluded or discovered by different experiments, 
except for the LHC discovery contour, where the region above the 
contour would be discovered.}
\label{mh_beta}
\end{figure}

To further understand the difference in the Higgs search reaches  due
to $\tan\beta$, we present  the mass contours for $m_{h}$ in $m_{\rm
aux}$ -- $\tan\beta$ plane, with  $m_{0}=1$ TeV (Figure~\ref{mh_beta}).
For $\mu>0$, we only study $\tan\beta<30$ due to the constraints
from $b\rightarrow{s}\gamma$.  LEP at $\sqrt{s}=200$ GeV can only
exclude a very small  corner up to $\tan\beta=6$ ($\tan\beta=7$) for
negative (positive) $\mu$.  We have to wait until the Tevatron starts
its  RUN II to fully exclude the entire region if no Higgs is
found. At the  same time, to discover a Higgs, it is only possible 
for LEP 200 when $\tan\beta<5-6$ or 
Tevatron at  10 ${\rm fb}^{-1}$ when $\tan\beta<8-10$.
Only  Tevatron operating at an integrated luminosity
larger than  20 ${\rm fb}^{-1}$ can cover the entire space.  LHC with 
luminosity 30 ${\rm fb}^{-1}$ can not explore the low $\tan\beta$ region.
With higher luminosity or in combination with other search channels, 
the entire region can be studied.  

\section{Conclusions and Discussions}
\label{conclusion}
The minimal AMSB scenario is a well defined low energy effective 
field theory with
all the SUSY spectrum determined by the four parameters :  $m_{\rm
aux}, m_{0}, \tan\beta$ and sign($\mu$).   The sign of $\mu$ changes
the results of $m_{A}$ at a large $\tan\beta$ and  $m_h$ at a small
$\tan\beta$.  The Higgs sector in this  model
is simplified because $m_{A}\gg{m}_{Z}$.  The lightest CP-even
Higgs is standard-model-like with mass in the range of 92 $-$ 118 GeV.
Current and future experiments in LEP, Tevatron and LHC have great
potential  to constrain this scenario.  If no Higgs is found in LEP
200, the low $\tan\beta$ region will be excluded.  Furthermore, if no
Higgs is found after Tevatron RUN  II at 2 ${\rm fb}^{-1}$, the AMSB
scenario will be totally excluded.  However,  if AMSB is true, a light
Higgs will first be detected at 5$\sigma$ significance level 
for low $\tan\beta$ case at LEP 200.  Tevatron running at luminosity of
10 ${\rm fb}^{-1}$ will discover a light Higgs if $\tan\beta$ is
small. Once Tevatron reaches its RUN III at 20 ${\rm fb}^{-1}$, or LHC
turns on, almost the entire parameter space of the 
AMSB scenario will be explored
and a signal of Higgs would  not escape detection. 

Comparing with the reaches of LEP, Tevatron and LHC for Higgs searches
in MSSM, mSUGRA and GMSB \cite{mssmsearch, othersearch}, we can see 
that the Higgs sector in AMSB scenario corresponds to the heavy $m_A$
case in MSSM.   Similar to mSUGRA and GMSB, low 
$\tan\beta$ region can be probed at LEP or at Tevatron running with lower 
luminosity, while the whole parameter space would be explored at Tevatron 
with higher luminosity and at LHC. 

As mentioned above, here we choose a common mass $m_{0}$ for all the
super  scalar masses in our analysis.  This is a simplified
assumption  and is not  generically true.  As the lepton-slepton sector
and first two generations  do not contribute much to the Higgs  sector
(due to the small Yukawa couplings), our results will remain true
if the sleptons and the first two generation squarks have  totally
different positive contributions.   The upper-left shaded  region in
$m_{\rm aux}$ -- $m_{0}$ spaces that was  excluded before because of the
negative slepton masses can now be  restored (the continuity of the
mass contours are also shown).  Furthermore, the positive contributions
to $m_{H_{d}}$ and $m_{D}$ can be different from those to
$m_{H_{u}}$ and $m_{Q,U}$, while $m_h$ remains almost the
same. This is because for $m_{A}\gg{m_{Z}}$, $m_{h}$  is determined by
$\tan\beta$ and the radiative sector from the stop sector,  which is
related to $m_{Q}$, $m_{U}$ and at small $\tan\beta$, to $\mu$.  Furthermore,
$m_{D}$ affects $\mu$ through $m_{H_{d}}$, as the running of RGEs are coupled
(see Eq.~\ref{running}).
However, the bottom Yukawa coupling at small $\tan\beta$ is small,  which
partly weakens the influence of $m_{D}$ on $m_{H_{d}}$, and further on 
$\mu$ and
$m_{h}$.  The effect of $m_{H_{d}}$ to $\mu$ is small  due to the
$1/(\tan^2\beta-1)$ factor in front of $m_{H_{d}}^2$ in determining $\mu$:
\begin{equation}
\mu^2=\frac{m_{H_{d}}^2-m_{H_{u}}^2\tan^2\beta}{\tan^2\beta-1}
-\frac{1}{2}m_{Z}^2.
\end{equation}
Nevertheless, $m_{A}$ would be changed if $m_{H_{d}}$
and $m_{D}$ take a different value of $m_0$.  For the large $\tan\beta$ case,
we can further change $m_{0}$ in $m_{H_{u}}$ while keeping $m_{h}$ to be
almost the same.  In this case, $m_{H_{u}}$ comes into effect 
through  $\mu$, whose effect
is small for large $\tan\beta$.  Thus, our conclusions will remain 
true if we take $m_{0}$ to be  the common mass correction only to
$m_{Q}$, $m_{U}$ and for  small  $\tan\beta$ also to $m_{H_{u}}$.
Even if all the additional positive  contributions to  the scalar
masses are different, our  analysis can still be  applied
as the only changes are in the boundary conditions.  Of
course   in this case, we have more parameters and the reaches in the
parameter  spaces from Higgs searches  become a bit more
complicated.

\noindent{\bf Acknowledgments:} It is a great pleasure to thank Lisa
Randall for enlightening advice and discussions.  We also thanks
Marcela Carena,  Howard Haber, Takeo Moroi and Jonathan Feng for
helpful correspondence and  conversations.  This work is supported by
the Department of Energy under  cooperative agreement
DF-FC02-94ER40818.

\end{document}